\documentstyle[11pt,newpasp,twoside,epsf]{article}

\begin{document}

\title{Model atmosphere analyses of post-common envelope stars}

\author{Mark T. Rushton, Barry Smalley, Richard N. Ogley, 
Janet H. Wood}
\affil{Astrophysics Group, School of Chemistry \& Physics, Keele University,
Staffordshire, ST5 5BG, United Kingdom}
\author{Peter H. Hauschildt}
\affil{Department of Physics \& Astronomy and Center for Simulation Physics,
The University of Georgia,
Athens, GA 30602-2451, USA}
\affil{Hamburger Sternwarte, Gojenbergsweg 112, 21029 Hamburg, Germany}
\author{James N. Bleach}
\affil{
Kyoto University, Yukawa Institute of Theoretical Physics,
Kitashirakawa, Sakyo-ku, Kyoto 606-8502, Japan}

\begin{abstract}

Understanding post-common-envelope evolution is important in the studies of
close-binary systems. The majority of all interacting binaries with a compact
object in their system (e.g. cataclysmic variables, X-ray binaries) are thought
to have passed through a common-envelope (CE) phase.  Current models of
interacting binaries all assume, as a starting point, that there is no
significant modification of the secondary star compared with a normal star of
similar spectral type.  The extent to which the CE significantly alters the
composition of the secondary star has yet to be determined.

We are studying the M-type secondary in the pre-CV system EG UMa, in order to
determine its atmospheric parameters by comparison with synthetic spectra
generated using {\sc phoenix}. Absorption lines due to several elements have been
used in investigating the stellar parameters of effective temperature, gravity
and over elemental abundance. In addition, we are searching for anomalies due
to s-process elements (such as Ba, Sr, Rb, Y), which may have been deposited in
the atmosphere during the CE stage, and prove that CE evolution has occurred.

EG UMa displays strong YO absorption bands, which are normally associated with
giants and S stars.  These bands may have been formed as a result of an
Y-abundance enhancement introduced by the CE, which would be consistent with
the observed possible Rb and Sr enrichment.

\end{abstract}

\section{Introduction}

Pre-Cataclysmic Variables (pre-CVs) are objects consisting of a white dwarf and
a late, low-mass main-sequence star.  The distinguishing feature of these
objects is that they have already undergone a phase of common-envelope (CE)
evolution, and are sometimes also called post-common-envelope binaries (PCEBs).

Initially the binary separation is large, however the CE phase allows a large
amount of angular momentum to be lost in the system, and the binary separation
shrinks.  Pre-CV or PCEB stars have not yet reached the evolutionary stage of
significant mass-transfer due to Roche-lobe overflow when they would enter the
cataclysmic-variable stage. Pre-CVs are therefore excellent objects for
studying the initial evolutionary stages as observations are not confused by
discs or hotspots.

EG UMa was initially identified as a DA white dwarf (Stephenson 1960) and later
a binary companion with emission lines was discovered (Greenstein 1965). The
following system parameters have been determined (see Bleach et al. 2000 for
details and references): The primary is a cool white dwarf at around 13,000 K,
with a mass of 0.64 M$_{\odot}$ and a radius of 0.013 R$_{\odot}$, while the
secondary is an M4--5 dwarf with a mass of 0.42 M$_{\odot}$ and a radius of
0.45 R$_{\odot}$. The orbital period of the system is around 16 hours (Lanning
1982).

However, previous work has indicated that the M dwarf might have a larger
radius and mass than that for a main-sequence star of similar spectral type
(Bleach et al. 2000). We can use high-resolution spectroscopy to obtain
estimates of the gravity in the M dwarf, which would give some insight into the
mass and radius inconsistencies.

\section{Observations}

Echelle spectroscopy was taken with the Utrecht Echelle Spectrograph (UES) on
the 4.2-metre William Herschel Telescope (WHT) during two nights in March 1998.
A total of 13 spectra of EG UMa together with 9 normal M dwarf stars, ranging
in spectral types between M0--6, were observed. Wavelength ranges from
$\sim$4250 -- 9000 \AA\ covering, much of the red region where the red dwarf is
dominant, and also included the Na I doublet at 8170--8220 \AA\ and the Ca II
triplet around 8540 \AA.

The data were reduced using the {\sc echomop} (Mills, Webb \& Clayton
1997) and the {\sc figaro} (Shortridge et al. 1999) packages. The blaze function
was removed from the spectra by normalizing with a polynomial fit to the
observed continuum (Bleach et al. 2002a). A spectral atlas of EG UMa has been
produced and is discussed in Bleach et al. (2002b).

\section{Synthetic Spectra}

We have generated a grid of high-resolution (0.05\AA\ steps) solar-composition
synthetic spectra using {\sc phoenix} code and the NextGen grid of model
atmospheres (Hauschildt et al. 1999 and references therein), covering the
$T_{\rm eff}$ range 2500--3900~K in steps of 100~K and the $\log g$ range
3.5--5.5 in steps of 0.5 dex. The synthetic spectra were rotationally broadened
to match the observations: EG UMa has $v \sin i$ = 27.8 kms$^{-1}$ (Bleach et
al. 2002a), while the standards have values less than 3 kms$^{-1}$.

\section{Atmospheric Parameters of EG UMa}

\begin{table}
\caption{Atmospheric parameters of EG UMa and normal M dwarfs, obtained from
line-profile fitting. The fits were made to solar-composition synthetic spectra.
Spectral types are based on those given in Bleach (2001).}
\label{stellar-params}
\begin{center}
\begin{tabular}{llll} \tableline
Star     & Sp. Type & $T_{\rm eff}$  & $\log g$      \\ \tableline
EG UMa   & M4.0--5.0 & 3300 $\pm$ 100 & 5.5 $\pm$ 0.5 \\
Gl\,486  & M3.5--4.0 & 3400 $\pm$ 100 & 5.0 $\pm$ 0.5 \\
Gl\,447  & M4.0--4.5 & 3300 $\pm$ 100 & 5.0 $\pm$ 0.5 \\
Gl\,548A & M0.0--1.0 & 3900 $\pm$ 100 & 4.5 $\pm$ 1.0 \\
Gl\,548B & M0.5--2.0 & 3900 $\pm$ 100 & 4.5 $\pm$ 1.0 \\
Gl\,403  & M3.0--3.5 & 3350 $\pm$ ~50 & 5.0 $\pm$ 0.5 \\
Gl\,436  & M2.5--3.5 & 3400 $\pm$ 100 & 5.0 $\pm$ 0.5 \\
Gl\,402  & M4.0--5.0 & 3300 $\pm$ 100 & 5.0 $\pm$ 0.5 \\
Gl\,406  & M5.5--6.0 & 2700 $\pm$ 200 & 5.0 $\pm$ 0.5 \\
Gl\,699  & M4.0--5.0 & 3200 $\pm$ 100 & 5.0 $\pm$ 0.5 \\
\tableline
\end{tabular}
\end{center}
\end{table}

To constrain the effective temperature and surface gravity of our programme
stars, line-profile fitting to prominent lines in the spectra, namely K, Na I,
Rb, TiO, Ti and Fe, was performed. Table \ref{stellar-params} gives the
best-fitting parameters obtained from the line-profiles for all objects in our
programme. Overall the K, TiO, Ti and Fe lines gave consistent results,
indicating that this multi-line methods appears to give reliable atmospheric
parameters. For example, our parameters for Gl\,406 agree very well with those
obtained by Basri et al. (2000).

\begin{figure}[ht]
\plotone{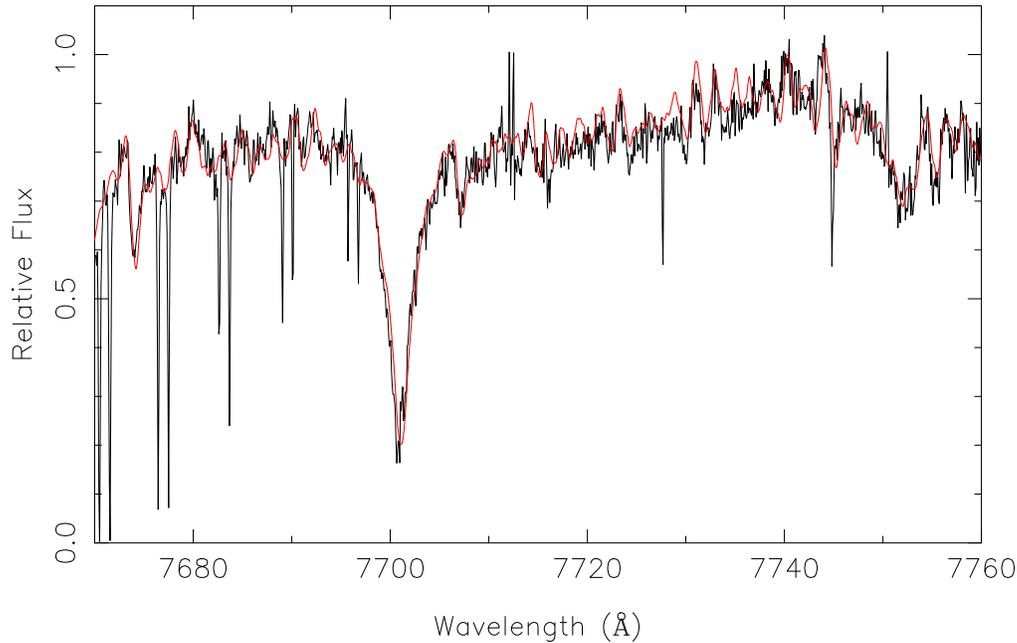}
\caption{Best broadened synthetic spectral fit for $T_{\rm eff}$ =
3300 K, log $g$ = 5.5 (red line).  The observed K 7699 \AA\ line is
shown in black. Telluric O$_2$ lines are clearly visible shortward of the
potassium line.}
\label{K-line}
\end{figure}

EG UMa is fitted more consistently by a higher surface gravity than normal
dwarf M stars, as shown by the K (7699 \AA) line in Fig.~\ref{K-line}. Good
fits to the core and wings are obtained, as well as to adjacent spectral
features, by using the $T_{\rm eff}$ = 3300 K and $\log g$ = 5.5
solar-composition synthesis. The higher surface gravity indicates either a
lower radius or larger mass (or both) than a main-sequence star of the same
spectral type. This is somewhat at odds with the Bleach et al. (2000)
conclusion that the radius is larger than a main-sequence star. The parameters
given by Bleach et al. (2000) would indicate a value of $\log g = 4.75 \pm
0.10$, which is closer to the values found for the normal M dwarfs.

The equivalent widths of the Na I doublet (8170--8220\AA) in EG UMa gives an
associated spectral range of M4--5, which is in agreement with the previous
work of Bleach et al. (2000). Inspection of Table~\ref{stellar-params} reveals
that this spectral type is consistent with the $T_{\rm eff}$ obtained from the
spectral fits.

Minor discrepancies do exist, however, between the synthetic spectra and the
observations. The Na D lines (5896, 5890\AA) and Ca II triplet (8498, 8542,
8662\AA) are somewhat stronger in the syntheses. The line broadening for the
strongest lines is not too well known. For example, using the Van der Waals
broadening calculations based on the Anstee, Barklem and O'Mara (ABO)
cross-section data give different line strengths (Barklem, Anstee \& O'Mara
1998, Barklem \& O'Mara 1998, Barklem, Piskunov \& O'Mara 2000). In addition,
possible non-LTE effects may also be important. Nevertheless, the overall fits
to both EG UMa and the normal M stars are certainly very impressive. In EG UMa,
however, the Na and Ca lines are affected by emission (Bleach et al. 2000,
2002b) and as such were not used in the determination of the atmospheric
parameters.

Overall, the M star in EG UMa appears to be relatively normal and well fitted
by a solar-composition synthetic spectrum. This indicates that the gross
properties of the star might not be too far from that of a normal non-binary M
dwarf. However, several spectral features revealed that EG UMa is not a
totally normal star.

\section{Yttrium oxide Bands}

\begin{figure}[ht]
\plotone{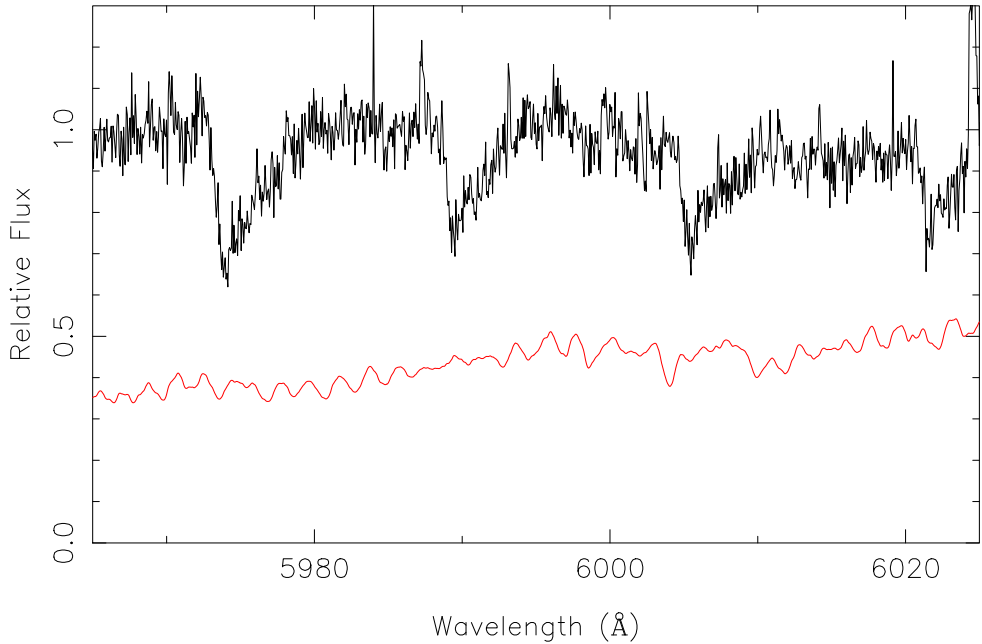}
\caption{Strong YO absorption bands in EG UMa. The black line shows
the observed data, and the red line is a synthetic spectrum with
$T_{\rm eff}$ 3300 K and $\log g$ = 5.5. One can see that there is no YO
absorption in this synthesis with solar-abundances.}
\label{YO-bands}
\end{figure}
\begin{figure}[ht]
\plotone{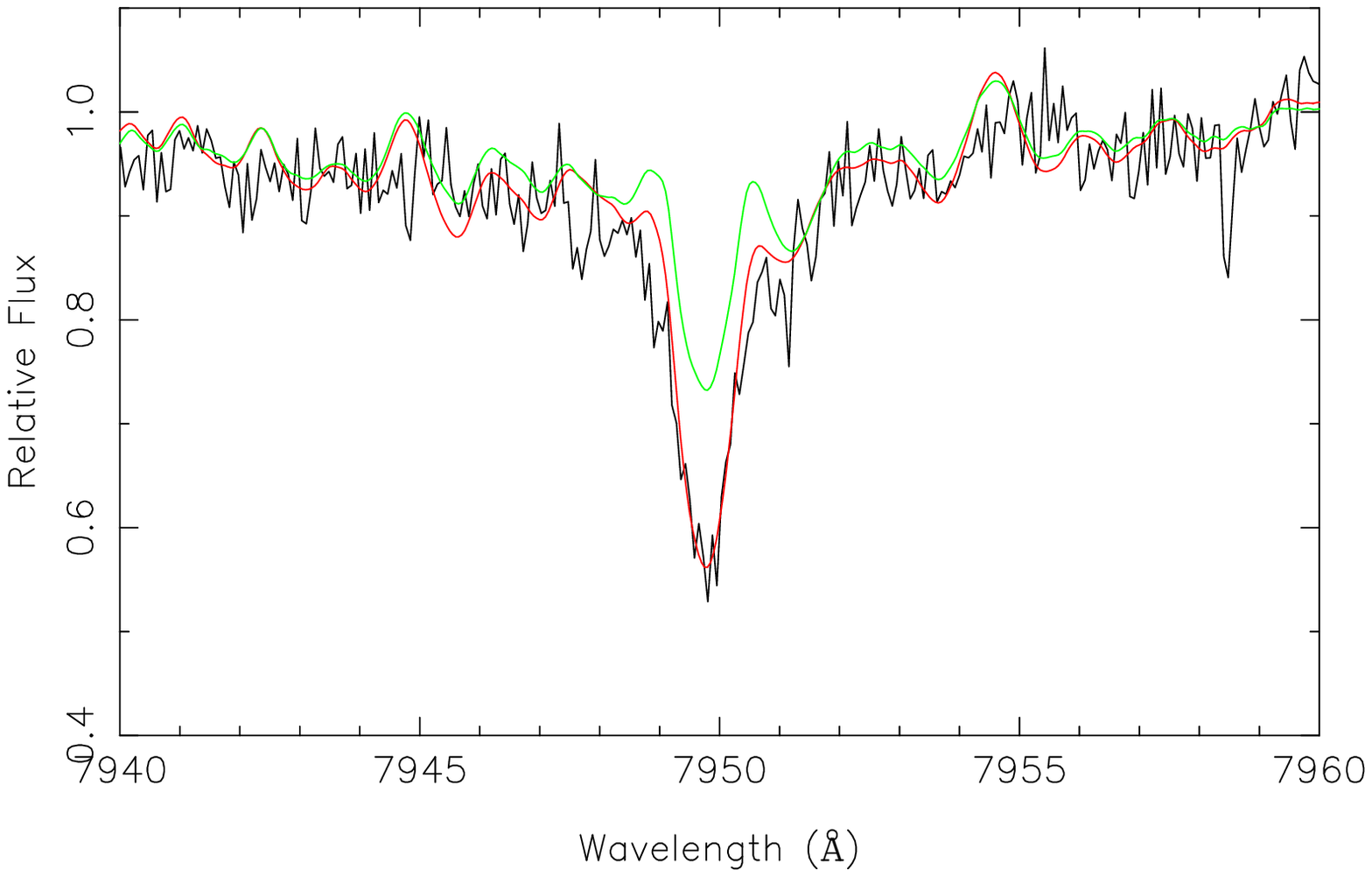}
\caption{The Rb line (in black) together with synthetic
spectral fits for $T_{\rm eff}$~=~3100~K, log $g$ = 5.5
(red), and $T_{\rm eff}$ = 3300 K, log $g$ = 5.5
(green).  The Rb line could represent an over-abundance with
respect to the (green) 3300 K model, which is the best overall fit to EG UMa.}
\label{Rb-line}
\end{figure}

In EG UMa there are several prominent molecular bands which are not observed in
the standard stars.  These bands are present in both nights, and in adjacent
orders, and follow the radial velocity shifts of the secondary star around the
binary orbit, so they are believed to be real stellar features. They were found
at 5972, 5988, 6004, 6020, and 6132 \AA, and are due to Yttrium Oxide (YO)
(Bernard et al. 1979, Bernard \& Gravina 1983). The first four bands are shown
in Fig.~\ref{YO-bands}.

The discovery of YO bands in the spectrum of EG UMa was totally unexpected and
potentially very significant. YO bands are usually only found in S and the
latest M stars (Keenan 1966, Scalo 1974, Wyckoff \& Clegg 1978, Ake 1979).  EG UMa is not an S star because it
displays intense TiO absorption.  YO typically appears at much lower gravities
due to changes in the chemistry from high to low pressures. Therefore, it is
found in giants, but not usually in dwarfs.

Resolutions to this paradox are: star spots (EG UMa is an active star, Bleach
et al. 2002b); a giant secondary; an extended atmosphere; or a distorted M
dwarf with a low gravity region. Star spots are unlikely because the bandheads
would show orbital phase variation, which is not observed. However, we cannot
rule out the possibility that the stellar surface is inhomogeneous. The M dwarf
cannot be a giant because log $g$ clearly is outside the range for giants, and
also the star would have to remain well within its Roche lobe or it would
become a cataclysmic variable. The system parameters indicate that EG UMa is
not tidally distorted, and its gravity does not change by more than 0.02 dex
from equator to pole (Maxted, private communication).

Alternatively, YO could represent an abundance enhancement of Y.  This would be
expected in a post-CE binary because Y, along with Ba, Sr and Rb, are all
s-process elements, which could have been accreted during the CE stage.
Support for this scenario is given by a strong Rb enhancement
(Figure~\ref{Rb-line}) and a possible enhancement of Sr. However, the Ba lines
appear to be normal.

\section{Conclusions}

Line-profile fitting has been used to determine the atmospheric parameters of
the secondary star in the EG UMa binary system. The best fitting solution gives
a surface temperature of the M star as 3300 $\pm$ 100 K and a surface gravity
of $\log g$ = 5.5. This value of $\log g$ is slightly higher than the average
for an early M dwarf.  EG UMa is unique among the stars in the sample by
displaying YO bands.  These may be explained by an abundance enhancement of Y,
which would be consistent with s-process enhancement from the common-envelope
phase of the binary system's evolution.  Similarly, such enrichment has been
inferred for Sr and Rb, but not for Ba. Further work is in progress to examine
the anomalies in more detail.

\end{document}